\documentclass[a4paper]{article}
\usepackage{graphicx} 
\usepackage{amsmath}
\usepackage{authblk}
\usepackage{bm}
\usepackage{algorithmic,algorithm}
\usepackage{url}
\newcommand*\samethanks[1][\value{footnote}]{\footnotemark[#1]}

\title{Reputation Games for Undirected Graphs} 

\author{
David Avis\thanks{School of Informatics, Kyoto University
and School of Computer Science, McGill University \texttt{avis@cs.mcgill.ca}}
\and
Kazuo Iwama\thanks{School of Informatics, Kyoto University}
\and
Daichi Paku\samethanks
}
\date{October 19, 2012}

\begin{document}

\newtheorem{theorem}{Theorem}[section]
\newtheorem{lemma}[theorem]{Lemma}
\newtheorem{proposition}[theorem]{Proposition}
\newtheorem{corollary}[theorem]{Corollary}

\newenvironment{proof}[1][Proof]{\begin{trivlist}
\item[\hskip \labelsep {\bfseries #1}]}{\end{trivlist}}
\newenvironment{definition}[1][Definition]{\begin{trivlist}
\item[\hskip \labelsep {\bfseries #1}]}{\end{trivlist}}
\newenvironment{example}[1][Example]{\begin{trivlist}
\item[\hskip \labelsep {\bfseries #1}]}{\end{trivlist}}
\newenvironment{remark}[1][Remark]{\begin{trivlist}
\item[\hskip \labelsep {\bfseries #1}]}{\end{trivlist}}

\newcommand{\qed}{\nobreak \ifvmode \relax \else
      \ifdim\lastskip<1.5em \hskip-\lastskip
      \hskip1.5em plus0em minus0.5em \fi \nobreak
      \vrule height0.75em width0.5em depth0.25em\fi}

\maketitle

\begin{abstract}
J. Hopcroft and D. Sheldon originally introduced network reputation games
to investigate the self-interested behavior of 
web authors who want to maximize their PageRank 
on a \emph{directed} web graph by 
choosing their outlinks in a game theoretic manner.
They give best response strategies for each player
and characterize properties of web graphs which are Nash equilibria.
In this paper we consider three different models for 
PageRank games on \emph{undirected} graphs such as certain social networks.
In undirected graphs players may delete links at will, but typically cannot
add links without the other player's permission.
In the \emph{deletion-model}
players are free to delete any of their bidirectional links but may not add links.
We study the problem of determining whether the given graph represents
a Nash equilibrium or not in this model.
We give an $O(n^{2})$ time algorithm for a tree,
and a parametric $O(2^{k}n^{4})$ time algorithm for general graphs,
where $k$ is the maximum vertex degree in any biconnected component of the graph.
In the \emph{request-delete-model} players are free to delete any bidirectional 
links and add any directed links,
since these additions can be done unilaterally and can
be viewed as requests for bidirected links. For this model we give an
$O(n^3)$ time algorithm for verifying Nash equilibria in trees.
Finally, in the \emph{add-delete-model} we allow a node to make arbitrary deletions 
and the addition of a single bidirectional
link if it would increase the page rank of the other player also.
In this model we give a parametric algorithm for verifying Nash equilibria in
general graphs and characterize so called $\alpha$-insensitive Nash Equilibria.
We also give a result showing a large class of graphs where there is an edge addition
that causes the PageRank of both of its endpoints to increase, suggesting
convergence towards complete subgraphs.
\end{abstract}

%\keywords{
%PageRank,
%Game theory,
%Nash equilibria,
%Fractional optimization
%}

\section{Introduction}

Introduced by Larry Page and Sergey Brin \cite{sergey}, the PageRank 
of a web page is
an important basis of the Google search engine and possibly one of the
most successful applications of a mathematical concept 
in the IT world.
PageRank is a value that is assigned to each web page according to the
stationary distribution of an $\alpha$-random walk on the web graph.
Here an $\alpha$-random walk is a random walk modified to make a
random jump with probability $\alpha$ at each step and a random jump
is a move to a node according to a given distribution vector $\bm{q}$.

Unlike rankings based on content such as keywords, tags,
etc., PageRank focuses solely on the hyperlink structure of the given
web graph.  Web links themselves possess strategic worth and
hence web authors often try to boost the PageRank of their web pages
by carefully choosing links to other pages.  Since these authors
behave strategically in a self-interested way, this is a typical
example of a non-cooperative game.  In fact, Hopcroft and Sheldon
recently introduced the PageRank game as a game theoretic model played
over a directed graph \cite{bib1}.  Each player is identified with a node, and a
strategy is a specification of a set of outlinks to other nodes.  The payoff
for each player is the PageRank value for their node which is
calculated on the resulting directed graph. The obvious goal of each
player is to maximize their payoff.

In \cite{bib1}, the authors proved a nice property of this game,
namely the best strategy of a player $v$ is to place her outlinks
to the nodes $u$ having largest potential value $\phi_{uv}$. The
potential $\phi_{uv}$ measures the probability of returning to
$v$ before the first jump and
does not depend on the outlinks from $v$ if
the other nodes do not change their outlinks.  Thus,
a simple greedy algorithm exists for deciding if a given graph is in
Nash equilibrium
and a nice characterization of
Nash equilibria graphs is possible.  Interestingly, it turns out that
such graphs representing Nash equilibria have very strong regularity
properties (see Section 
\ref{edgedeletion} for details).
The purpose of this paper is to study similar problems on undirected graphs.

{\bf Motivation.}
Social networks have become one of the defining paradigms of our time, with enormous
influence on how decisions are taken and events unfold. As with web graphs, content
by itself will rarely be enough to explain the dynamics of these networks.
The underlying graph structure itself surely plays a role in how new relations are formed
and old relations broken. In considering two major social networks, Facebook and
Twitter, a casual glance shows a radically different graph structure
in spite of the fact that they have a comparable number of similar users. Facebook, an
undirected graph, has few nodes with degree more than a 1000. Twitter, a directed graph,
has nodes (such as Kate Perry) with in-degree 28 million and out degree just 115.
A basic difference in the dynamics of the
two networks is edge addition, which requires the
approval of both nodes in an undirected graph but does not in a directed graph.
The ability to add and delete links instantly in a directed network allows for
an extremely rapid dynamically changing graph structure.
Anecdotal evidence points to a much more stable graph structure in Facebook which
apparently consists of large number of relatively small very dense subgraphs.

Another example of an undirected network is the graph of international
bilateral agreements between universities. 
We might consider PageRank as measuring how prestigious a university is,
and universities might only accept agreements if it increases their prestige. 
Finally we might consider the coauthorship graph, possibly one of the oldest
social networks, defined so that people could find their ``Erdos number''.
Here edge deletions are not permitted, but edge additions could conceivably
be influenced by the PageRank of the given nodes.

Our motivation is to build models for undirected graphs and to study
their dynamics. Our basic tool will be to adopt PageRank as a quantity that
users try to optimize. Under this assumption we will study how undirected networks
evolve, what networks in equilibrium look like, and contrast this to 
the case of directed networks. 
Whether users of these networks actually behave in this way
is beyond the scope of this paper.

\begin{figure}
\centering
\includegraphics[scale=1.0]{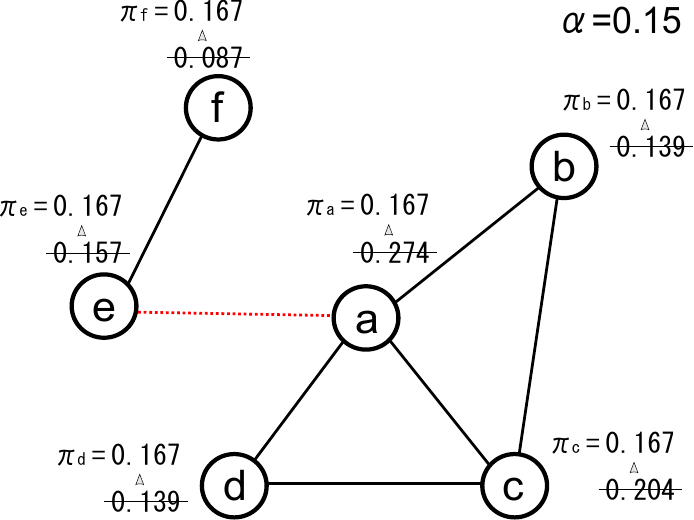}
    \caption{Edge deletion}
\label{deletion}
\end{figure}

{\bf Outline of the paper.}
We introduce three different models for PageRank games
on undirected graphs.
Our study mainly focuses on the \emph{deletion-model} 
(described in Section \ref{edgedeletion})
where a player cannot create a new link
but may unilaterally delete an existing one.
In the directed web graph model described above,
what $v$ intuitively does is
to cut its links to nodes $u$ having smaller $\phi_{uv}$ values, assuming
that $u$ will not delete its edge to $v$.  
In the deletion-model, if $v$
cuts its outlink to $u$ we assume that $u$ also cuts its outlink to $v$
either automatically, or as a form of revenge.  
In Figure \ref{deletion} deleting edge $ae$ increases $e$'s PageRank but decreases $a$'s.
So $e$ may unilaterally choose to do this. Note after the edge deletion, if $a$
proposed to reinstate the edge $e$ would refuse.

Unlike the directed graph model, in the undirected model a node cannot add
a new edge by acting unilaterally, 
but must seek the permission of the other node of the edge.
It turns out that the class of equilibria graphs in the deletion-model
is larger than in the directed model.
Unfortunately,
the nice property of the original model that $\phi_{uv}$ does
not depend on the outlinks from $v$, no longer holds. Hence the
greedy algorithm for the Nash decision problem does not work, either,
and there seems to be no obvious way of checking 
the equilibrium condition.

In Section \ref{trees} we give an $O(n^{2})$ time algorithm for
the case where the graph is a tree.
In Section \ref{general}
we gave a parametric algorithm for general graphs, where the parameter $k$ is
the maximum degree of any vertex in any biconnected component that
contains it. 
Biconnected components roughly correspond to local clusters of web pages,
where one could expect the parameter $k$ to be relatively
small. Nodes linking biconnected
clusters may have arbitrarily large degree without
changing the time complexity.
We give an $O(2^{k}n^{4})$ time
algorithm for general graphs.

Our second model 
is the \emph{request-delete-model} where a player can
unilaterally delete any existing edges and also can unilaterally
create any new directed outlinks but cannot create a new inlink.
In Section \ref{edgedeletion2},
we give an $O(n^{3})$ time algorithm for trees which determines
if the given graph is a Nash equilibrium in request-delete-models.
It draws on the algorithm in Section \ref{trees}.

\begin{figure}
\centering
\includegraphics[scale=1.00]{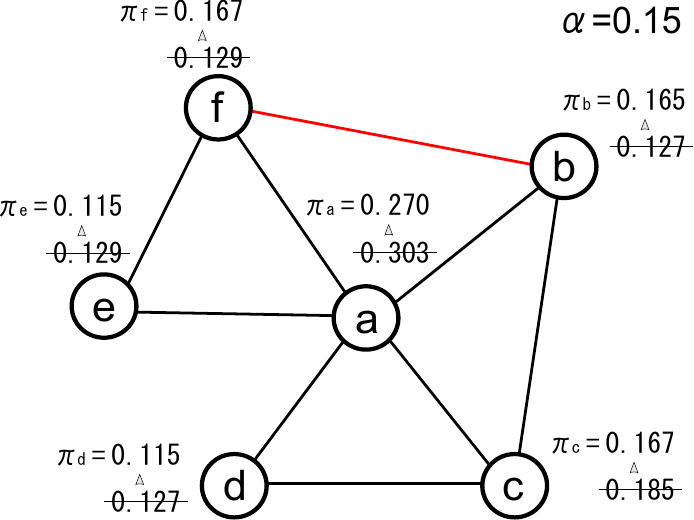}
    \caption{Edge addition}
\label{addition}

\end{figure}

The third model is called the \emph{add-delete-model} where a player can
delete any existing edges and also can add an undirected edge to
another player 
if the PageRank of both players are improved.
In Figure \ref{addition} adding edge $bf$ increases both $b$'s and $f$'s PageRank,
so both parties would accept this addition. Note the PageRank of all other players
decreases. 

While it may seem a great restriction to consider single edge
additions, recall that we are interested in Nash equilibria.
Multiple edge additions require the \emph{simultaneous} decisions of multiple players.
A player in this group cannot know the actions of the other
players and hence cannot predict the new graph structure.
Therefore it is not possible in general for a player to calculate whether or not
an edge addition would improve her PageRank. 
In Section \ref{adddel},
we give an $O(2^{2k}n^{5})$ time algorithm 
for general graphs which determines
if the given graph is a Nash equilibrium in add-delete-models.
It draws on the algorithm in Section \ref{general}.
We also give two structural type theorems. The first shows that the only
$\alpha$-insensitive equilibria are complete graphs. The second says that
in symmetric graphs edge-addition will occur. This gives some theoretical justification
for the anecdotal evidence cited earlier with respect to Facebook.

Our results begin with the study of trees. This is not because social networks
are likely to be trees but because 
the more complex parametric algorithms for general graphs are based on these
results. 

{\bf Related work.}  Although it focuses less on game theoretic
aspects, there is a large literature on optimal
linking strategies to maximize the PageRank of given nodes
for directed graphs.  On the positive side, 
Avrachenkov and Litvak \cite{avrachenkov} give a polynomial-time
algorithm
for maximizing the PageRank of a single node by selecting its outlinks.
Kerchove et. al. \cite{kerchove} extend this result to
maximizing the sum of the PageRank of a given set of nodes.
Csaji et. al. \cite{csaji} 
give a polynomial-time algorithm for maximizing the
PageRank of a single node with any given set of controllable links.  

On the negative side,  \cite{csaji} also shows that
the problem becomes NP-hard if some pairs of controllable links in the
set are mutually exclusive.
Olsen \cite{olsen1} proved that maximizing the minimum PageRank in
the given set of nodes is NP-hard if we are allowed to add $k$ new
links. He also 
proved that the problem is still NP-hard if we restrict the node set
to a single node and the $k$ links to only incoming ones to that
node \cite{olsen2} and gives a constant factor approximation algorithm
for this problem \cite{olsen3}. 
The question of whether there are $\alpha$-sensitive Nash equilibria was
recently affirmatively answered by Chen et. al. \cite{chen}.

This paper is an extended version of an earlier paper presented at ISAAC 2011 \cite{avis},
which only covered the edge deletion model.

\section{Preliminaries}
\label{preliminaries}

\subsection{PageRank values}

Initially we describe the Hopcroft-Sheldon directed graph model.
Let $D=(V, E')$ be a simple directed graph on node set $V$ and arc set $E'$,
and let $\bm{q}$ be a probability distribution on $V$.
Throughout the paper we let $n = |V|$ denote the number of nodes.
For $v \in V$, let $\Gamma(v)$ denote the set of $v$'s out-neighbours.
A random jump is a move to a node according to the distribution
vector $\bm{q}$ instead of using one of the outlinks of the current node.
An $\alpha$-random walk on $D$ is a random walk that is modified
to make a random jump with fixed probability $\alpha$ ($0 < \alpha < 1$) at each step.
The PageRank vector $\bm{\pi}$ over the $n$ vertices in $V$ is 
defined as the stationary
distribution of the $\alpha$-random walk.
We define the potential matrix $\Phi = (\phi_{uv})$ such that for
vertices $u, v \in V$, $\phi_{uv}$ is the probability that a random walk
that starts from $u$ visits $v$ before the first random jump ($\phi_{uv}=1$
if $u=v$), which can be written as 
\begin{equation}
\label{phi}
\phi_{uv}=\frac{1-\alpha}{| \Gamma(u) |}\sum_{i \in \Gamma(u)}\phi_{iv}.
\end{equation}
In order to calculate $\bm{\pi}$, we have the following equation \cite{bib1}:
\begin{equation}
\label{pi}
\pi_{v} = \alpha 
\frac{\sum_{u\in V}q_u\phi_{uv}}{1-\frac{(1-\alpha)}
{| \Gamma(v)|}\sum_{i \in \Gamma(v)}\phi_{iv}}.
\end{equation}
Chen et. al. proved that $\pi_{v}$ is continuous for $\alpha \in (0,1)$ \cite{chen}.

\subsection{Directed PageRank games}
\label{DPG}

In the PageRank games in \cite{bib1} the players are the
nodes $V$ of a directed graph $D$ and they attempt to
optimize their PageRank by strategic link placement.
A {\it strategy} for node $v$ is a set of outlinks.
An outcome is an arc set $E'$ for $D$ consisting of
the outlinks chosen by each player.
The payoff of each player is the value of PageRank which
is calculated on $D$.

We say a player $v$ is in {\it best response}, if $v$
takes a strategy which maximizes $v$'s PageRank in $D$.
A directed graph $D$ is a {\it Nash equilibrium} if the
set of outlinks for each node is a best response:
no player can increase her PageRank value by choosing different outlinks.
Several results for best response strategies and for
Nash equilibria were introduced in \cite{bib1}.
In particular they gave a characterization of $\alpha$-insensitive
Nash equilibria, which are graphs being Nash equilibria
for all values $0 < \alpha < 1$ of the jump parameter.

In this paper we study PageRank games for undirected graphs.
Let $G=(V,E)$ be an undirected graph on vertex set $V$
and edge set $E$. Define the directed graph $D=(V,E')$ on
the same vertex set $V$, where each edge $uv$ in $E$ gives
rise to two arcs $uv$ and $vu$ in $E'$. In our model, the payoff
of each player $v$ for the graph $G$ is the PageRank of $v$
in the corresponding directed graph $D$.

\section{Deletion Only Models}
\label{edgedeletion}
In this section, we study the deletion-model 
for undirected PageRank games,
where a player cannot unilaterally create
a bidirected link to another node, but it can delete a bidirectional link.
We consider the problem that determines whether the given graph is a 
Nash equilibrium in deletion-model or not.
In Section \ref{trees} we give a quadratic algorithm
for the special case when $G$ is a tree,
and in Section \ref{general}
we give an $O(2^{k}n^{4})$time algorithm for general graphs
where $k$ is the maximum vertex degree on any biconnected component on $G$.

In the deletion-model, we say that a player $v$ is in best response if 
$v$ cannot increase her PageRank by any deletion of her (bidirectional) links.
A Nash equilibrium is a graph for which every player is in best response.
We consider the following problem:

\begin{description}
\item[Input:] An undirected graph $G$, $\alpha$, $\bm{q}$.
\item[Output:] Is the input a Nash equilibrium? (yes/no)
\end{description}

An equivalent formulation is to decide whether no
player can increase her PageRank for the given input,
where she is only allowed to delete edges to her neighbours.
As for directed graphs, we let $\Gamma(v)$ denote
the neighbours of vertex $v$ in $G$.
A strategy for $v$ is to retain a subset
$E_v  \subseteq \Gamma(v)$ of neighbours and delete
edges to her other neighbours.
Let $\bm{x}$ be a $0/1$ vector of length $d_v = |\Gamma(v)|$,
which indicates $v$'s strategy.
Formally, if $i \in E_v$ then $x_i=1$, otherwise $x_i=0$,
for $i=1,...,d_v$.
Let $\phi_{uv}(\bm{x})$ denote the potential function (\ref{phi})
for the subgraph of $G$ formed
by deleting edges $(v,i), i \in \Gamma(v) - E_v$.
By (\ref{pi}) applied to the corresponding directed graph
$D$ the PageRank of $v$ can be written as
\begin{equation}
\label{pix}
\pi_{v}(\bm{x}) = 
\alpha\frac{\sum_{u \in V}q_{u}\phi_{uv}(\bm{x})}
           {1-(1-\alpha)\dfrac{\sum_{i \in \Gamma(v)}\phi_{iv}(\bm{x})x_i}
                             {\bm{1}^T \bm{x}}}
\end{equation}
where $\bm{x} \neq \bm{0}$.
Let $\bm{1}_m$ denote a vector of ones of length $m$.
Usually the length is clear by the context so for
simplicity we may drop the subscript $m$. 
If the input is a Nash equilibrium then $v$ is using a best response
and no edge deletions for $v$ will raise her PageRank.
Therefore $\pi_{v}(\bm{1}_{d_v}) \geq \pi_{v}(\bm{x})$ for any 0/1 vector $\bm{x}$.
The approach we will use to solve the problem described
in this section is to compute the maximum of $\pi_{v}(\bm{x})$
over all 0/1 vectors $\bm{x}$ of length $d_v$,
for each vertex $v$.

\begin{figure}[htbp]
  \begin{center}
    \includegraphics[width = 0.90\hsize]{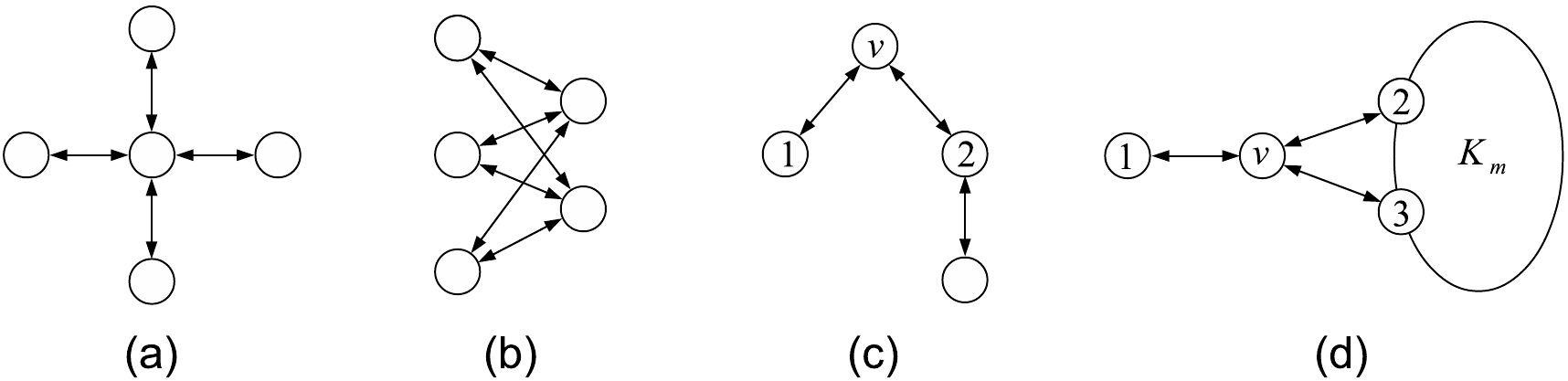}
    \caption{Examples}
    \label{graphs}
  \end{center}
\end{figure}

We give some examples in Figure \ref{graphs}.
Graphs (a), (b) are $\alpha$-insensitive 
Nash equilibria in directed PageRank games,
and are also a Nash equilibrium in deletion-models 
for any given $\alpha$ and $\bm{q}$.
Graph (c) is an example which is not a Nash equilibrium in directed games,
since $v$ can increase its PageRank
if we delete the arc from $v$ to $2$ which has less potential than $1$.
However (c) is a Nash equilibrium in deletion-models
for $\alpha=0.15$ and uniform distribution $\bm{q}$,
since if $v$ cuts its edge to $2$ then it decreases $v$'s PageRank.
Graph (d) is not a Nash equilibrium in both directed models and deletion-models
for $\alpha=0.15$ and uniform distribution $\bm{q}$,
where $K_{m}$ is a $m$-complete graph, $m = 8$ for example.
In this graph the potentials from $2$, $3$ to $v$ are much less than $1$,
so $v$ may try to cut the edge to $2$ or the edge to $3$,
but a single edge deletion decreases $v$'s PageRank.
Interestingly, the deletion of both edges leads to a greater PageRank for $v$.

\subsection{Trees}
\label{trees}
In this section, we study the problem
where the graph $G$ is a tree.
We prove the following theorem.
\begin{theorem}
\label{thmtree}
Given a tree $G$, jump probability $\alpha$ and distribution
$\bm{q}$, we can determine in $O(n^2)$ time whether $G$
is a Nash equilibrium in the deletion-model and if not give an improving strategy
for at least one player.
\end{theorem}
The remainder of this section is devoted to the proof of this theorem.

Let $v$ be a node in $G$, let $\Gamma(v)$ be the set of
neighbours of $v$, and let $d_v = |\Gamma(v)|$.
Consider any strategy for $v$ as described above
and let $\bm{x}$ be the 0/1 vector that represents it.
For $i \in \Gamma(v)$, let $N_i$ be the set of nodes
which are descendants of $i$ (including $i$ itself)
in the subtree of $G$ rooted at $v$.
For a node $u \in N_i$,
\begin{equation}
\phi_{uv}(\bm{x})= \phi_{uv}x_i,
\label{phitree}
\end{equation}
since potentials of all nodes in $N_i$ depend on only link
$(v, i)$ and the other links of $v$ do not affect these potentials.
This is because if $v$ cuts link $(v, i)$, all nodes in $N_i$
are disconnected from the other nodes in $G$.

Therefore (\ref{pix}) can be rewritten:
\begin{equation}
\label{tpiv}
\pi_{v}(\bm{x}) = \alpha\dfrac{\sum_{i\in \Gamma(v)}\sum_{u \in N_i}q_{u}\phi_{uv}x_i 
}{1-(1-\alpha)\dfrac{\sum_{i\in \Gamma(v)}\phi_{iv}x_{i}}{\bm{1}^T\bm{x}}}.
\end{equation}
Note that the potential matrix $\Phi$ on $G$ can be computed in $O(n^{2})$ time,
by using Gaussian elimination methods
for each column vector $(\Phi)_{v}$ defined by equation (\ref{phi}).
Since $G$ is a tree
we can apply elimination steps in post-order, where we consider $v$
to be the root of $G$.
There are at most $n$ forward eliminations and
backward substitutions because every node except $v$ has
only one parent.
Therefore it costs $O(n^{2})$ time to compute $\Phi$.

Let $a_i = \alpha \sum_{u \in N_i}q_{u}\phi_{uv}$ and
let $b_i = 1-(1-\alpha)\phi_{iv}$ for $i \in \Gamma(v)$.
Consider the fractional integer programming problem,
\[
P : \mbox{maximize} ~~~~
  \pi_{v}(\bm{x})=
    \bm{1}^T\bm{x}~
    \frac{\bm{a}^T\bm{x}}
         {\bm{b}^T\bm{x}},~~~\bm{x} \in \{0,1\}^n
\]
where $a_i \geq 0$ and $b_i \geq 0$ for $i\in \Gamma(v)$ are known constants.

In order to solve problem $P$, we fix the Hamming weight $l$ of $\bm{x}$
and we solve the following problem for each $l=1,...,d_v$:
\[
Q : \mbox{maximize} ~~~~
f(\bm{x})=
  \frac{\bm{a}^T\bm{x}}
       {\bm{b}^T\bm{x}}
  ~~~~~\mbox{subject to}
  ~~~~ \bm{1}^T\bm{x} = l.
\]
Problem $Q$ can be solved directly by Megiddo's method
in $O(n^{2}\log^{2}n)$ time \cite{megiddo},
and it can be also solved by Newton's Method 
in time $O(n^{2}\log n)$ \cite{radzik}.
However we are able to specialize Megiddo's method for
our problem to obtain an $O(n^{2})$ time algorithm.
Our approach initially follows the technique describe
in \cite{megiddo}.

\begin{figure}[htbp]
\begin{tabular}{lr}
  \begin{minipage}{0.45\hsize}
    \begin{center}
    \includegraphics[width = \hsize]{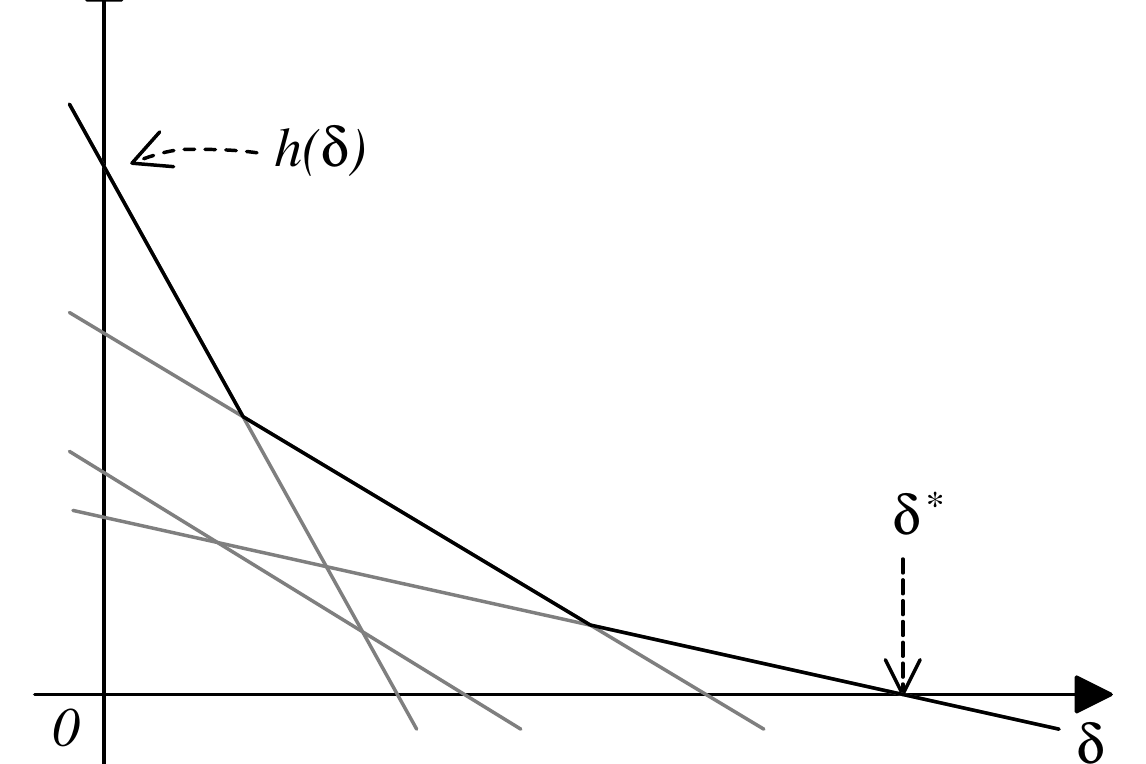}
    \caption{Solving $h(\delta)=0$}
    \label{lfco}
    \end{center}
  \end{minipage}
  \begin{minipage}{0.45\hsize}
    \begin{center}
    \includegraphics[width = \hsize]{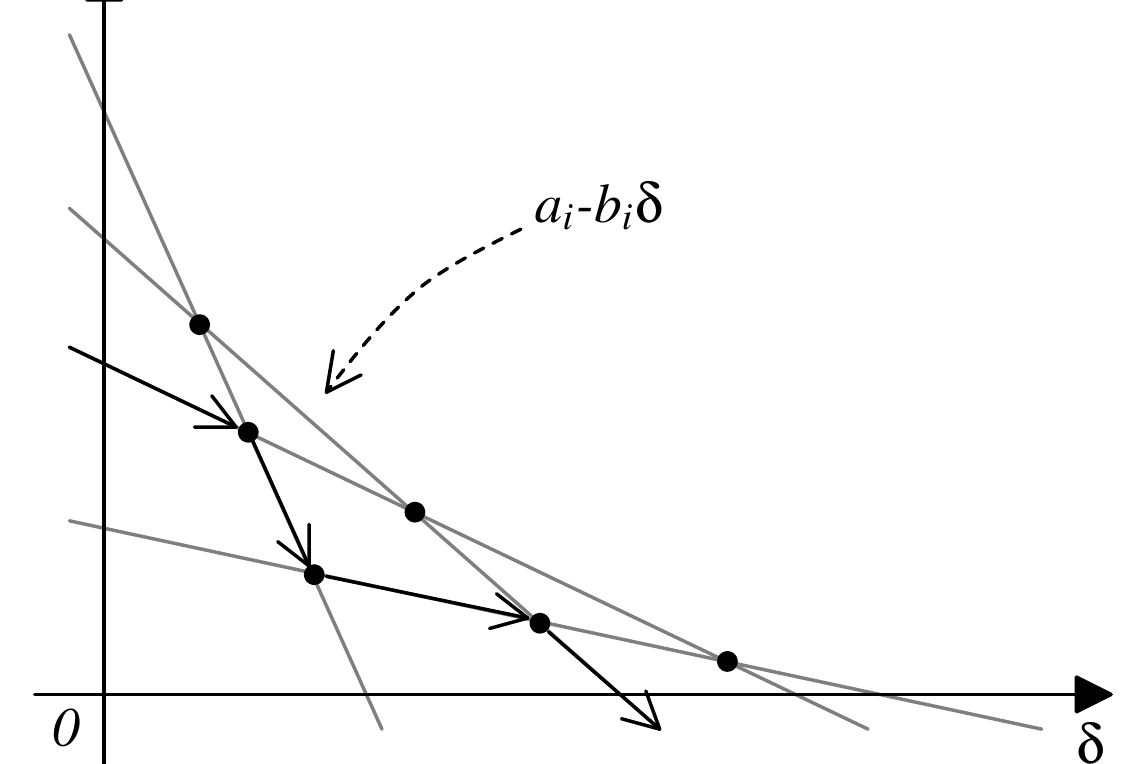}
    \caption{Path for $l=3$ on $H$}
    \label{linear}
    \end{center}
  \end{minipage}
\end{tabular}
\end{figure}

Since $\max \pi_{v}(\bm{x}) = \max_{l}l~f(\bm{x})$, we can solve
problem $P$ by solving problem $Q$ for each $l=1,...,d_v$.
Consider the following maximization problem for some fixed $\delta$.
\[
R : \mbox{maximize} ~~~~
g(\bm{x})=(\bm{a} - \bm{b} \delta )^T \bm{x}
~~~~~\mbox{subject to}
~~~~ \bm{1}^T \bm{x} = l.
\]
Let $c_i=a_i-b_i\delta,  i \in \Gamma(v)$ and let
$S(\delta)$ be the decreasing sequence of indices ordered by the values of $c_i$.
Problem $R$ is easily solved by choosing the first $l$ indices in $S(\delta)$.
Let $h(\delta)$ be the optimal value of problem $R$ for a given $\delta$,
that is, $h(\delta) = \max \{ g(\bm{x}) : \bm{1}^T\bm{x} = l \}$.
When $h(\delta) = 0$, then $\delta$ is equal to $\delta^{*}$
which is the optimal value of problem $Q$.
On the other hand, if $h(\delta) > 0$ then $\delta < \delta^{*}$,
and if $h(\delta) < 0$ then $\delta > \delta^{*}$,
i. e., a root of $h$ (see Figure \ref{lfco}).
The task in solving problem $Q$ is to find the value
$\delta^{*}$ for which $h(\delta^{*}) = 0$
by some tests on $\delta$.
The key point is how many values of $\delta$ have to
be tested for finding $\delta^{*}$.
Since the optimal solution for $R$ can change only when
the order of $S(\delta)$ changes,
we only have to test $\delta$ at intersection values of
lines $\{ c_i = a_i - b_i \delta : i\in \Gamma(v) \}$.
Using some results from computational geometry, we are able
to do this efficiently.
First we compute the line arrangement of lines
$\{ c_i = a_i - b_i \delta : i\in \Gamma(v) \}$.
Namely we define the planar graph $H$ which is formed
by subdivision of the plane induced by these lines.
Then we look all edges in $H$ as directed according to
the positive direction of $\delta$.

For each $l$ we test values of $\delta$ at the change
point of the $l$th entry in $S(\delta)$,
which are the nodes in $H$ lying on the $l$th layer
of the arrangement.
An example with $l=3$ is shown in Figure \ref{linear}.

We summarize the algorithm for solving $P$.
First compute constants $a_i$, $b_i$, for $i \in \Gamma(v)$,
and line arrangement of lines $\{ c_i = a_i - b_i \delta : i\in \Gamma(v) \}$.
Compute $S(0)$ by sorting the values of $a_{i}$.
For $l=1, ..., d_v$, do following steps:
let $x_{i} = 1$ if $i$ is within $l$th entry of $S(0)$,
and otherwise $x_{i} = 0$.
Compute $A = \bm{a}^{T}\bm{x}$ and $B = \bm{b}^{T}\bm{x}$.
From the starting edge, which is $l$th edge from the top,
follow the $l$-th layer as follows:
When we follow the edge on the line $c_{i}=a_{i}-b_{i}\delta$,
and visit the node which is intersection of $c_{i}=a_{i}-b_{i}\delta$
and $c_{j}=a_{j}-b_{j}\delta$,
let $A \leftarrow A - a_{i} + a_{j}$ and $B \leftarrow B - b_{i} + b_{j}$.
If $A - B\delta < 0$ then output $\bm{x}$ as the solution,
otherwise let $x_{i} \leftarrow 0$ and $x_{j} \leftarrow 1$,
and go to next node by following the edge on the line $c_{j}=a_{j}-b_{j}\delta$.

Finally, we analyze the running time of the algorithm.
Computing the line arrangement takes $O(d_{v}^{2})$ time,
which is done by the
incremental method or topological sort algorithms
(see Edelsbrunner \cite{edels}).
Since the number of nodes in $H$ is $O(d_{v}^{2})$ and
each of them is visited twice, 
we can find $\delta^{*}$ for each $l$ in at most $O(d_{v}^{2})$ time.
Therefore this algorithm solves problem $P$ in $O(d_{v}^{2})$ time.
Note that if $v$ is not in best response the solution to
$P$ gives an improving strategy for $v$.

As we have seen
we can test whether a node $v$ is in best response in $O(d_{v}^{2})$ time.
Because $G$ is a tree we have
$\sum_{v \in V}d_{v} = 2(n-1),$
and we have
$\sum_{v \in V}d_{v}^{2} < \left(\sum_{v\in V}d_{v} \right)^{2} = 4(n-1)^{2}.$
Therefore we can test whether all nodes are in best response in $O(n^2)$ time.
This concludes the proof of Theorem \ref{thmtree}.

\subsection{General Graphs}
\label{general}

In this section we give a parametric algorithm for general
connected graphs based on a parameter $k=k(G)$ defined
as follows.
If $G=(V,E)$ is a tree we set $k=1$ else $k$ is the maximum
vertex degree in any biconnected component of $G$.
Note that $k(G)$ can be computed in $O(|E|)$ time by decomposing
$G$ into its biconnected components and by finding
the maximum vertex degree in every such component.
Note that graphs can have a large maximum vertex degree but
small parameter $k$.
This would occur whenever the large degree vertices were cut vertices.
In a network setting the biconnected components could represent
small groups of well connected web pages with relatively few links
per page. These groups would be linked together by a few pages
containing many more links. 
We prove the following theorem.
\begin{theorem}
\label{genthm}
Given a graph $G$ with $k=k(G)$, jump probability
$\alpha$ and distribution $\bm{q}$, 
in $O(2^k n^4)$ time we can
determine if this is a Nash Equilibrium in the deletion-model and if not
give an improving strategy for at least one player. 
\end{theorem}
The remainder of this section is devoted to the proof of this theorem.

Let $v$ be a node in $G$ and
let $\{ C_1, C_2, ..., C_d \}$ be the set of connected components
in the subgraph that is induced by deletion of $v$ from $G$.
It follows from the definition of $k=k(G)$ that
$v$ has at most $k$ links to $C_{i}$ for $i=1,..., d$.
Let $U_{i} = \{ u : u \in C_{i}, u \in \Gamma(v) \}$ indicate
the set of $v$'s neighbours in $C_i$, 
for $i=1,...,d$. We have $|U_{i}| \leq k$ by the definition of $k$.
Consider any strategy for $v$, as described in Section \ref{edgedeletion}
and let $\bm{x}$ be the 0/1 vector of length $d_v = |\Gamma(v)|$ that represents it.
We write $\bm{x}=(x^1,x^2,...,x^d)$ as the concatenation of the 0/1
vectors $x^i$ representing the strategy
restricted to the component $C_i$, $i=1,...,d$.
Then, for $u \in V$, if $u \in C_i$ for $i=1,...,d$,
the potential of $u$ is written as follows:
\begin{equation}
\phi_{uv}(\bm{x})=
  \sum_{S \subseteq U_{i}}
    \phi^{S}_{uv}
    \prod_{s \in S}x^i_{s}
    \prod_{s \in U_{v}- S}(1-x^i_{s})
\label{phieq}
\end{equation}
where $\phi^{S}_{uv}$ are the potentials from $u$ for the subgraph of $G$ formed
by deleting edges $vu, u \in \Gamma(v)-S$.
This is because potentials to $v$ from all nodes
in $C_i$ depend only on links to $U_{i}$ and never
depend on other links.
To compute each column vector $(\Phi)^{S}_{v}$ for
$S \subseteq U_{i}$ for $i = 1, ..., d$,
we solve the linear systems defined on (\ref{phi})
by Gaussian elimination method in $O(2^{k}n^{3})$ time.

We have the formula for PageRank of $v$ as
\begin{equation}
\label{pixgen}
\pi_{v}(\bm{x}) = 
  \bm{1}^T\bm{x}
    \dfrac{\sum_{i}\sum_{S \subseteq U_{i}} a_{S}\prod_{s \in S}x^i_{s}}
          {\sum_{i}\sum_{S \subseteq U_{i}} b_{S}\prod_{s \in S}x^i_{s}}.
\end{equation}
where $a_{S}$ and $b_{S}$ be constants such that
\begin{eqnarray}
\label{generala}
a_{S} &=&
  \sum_{u \in C_{i}: S \subseteq C_{i}} \alpha q_{u}
  \sum_{T\subseteq S}(-1)^{|S|-|T|}\phi^{T}_{uv} \\
\label{generalb}
b_{S} &=&
\left\{
\begin{array}{ll}
  1-(1-\alpha)\phi^{S}_{uv} 
    & ~~~S = \{ u \}\\
  -(1-\alpha)\sum_{u\in S}\sum_{T \subseteq S}(-1)^{|S|-|T|}\phi_{uv}^{T}
    & ~~~\mbox{otherwise.}
\end{array} 
\right .
\end{eqnarray}
Note that $a_{S} \geq 0$ for all $S$ 
and the denominator of $\pi_{v}(\bm{x})$ is always positive.

In order to determine whether $v$ maximizes its PageRank,
consider the following fractional integer programming problem:
\[
P : \mbox{maximize} ~~~~ 
 \pi_{v}(\bm{x}), 
~~~ x^i_s \in \{ 0, 1\}
~~ \mbox{for}~ s\in U_i,~i=0,...,d.
\]
The method for problem $P$ is the similar to that used in Section \ref{trees}.
We fix the Hamming weight of $\bm{x}$ as $\bm{1}^T\bm{x}=l$,
and consider the following fractional integer programming problem:
\[
Q : \mbox{maximize} ~~~~ 
f(\bm{x})= 
  \dfrac{\sum_{i}\sum_{S \subseteq U_i} a_{S}\prod_{s \in S}x^i_s}
        {\sum_{i}\sum_{S \subseteq U_i} b_{S}\prod_{s \in S}x^i_s}
~~~~~\mbox{subject to} 
~~~~ \bm{1}^T \bm{x} = l.
\]
Since $\max \pi_{v}(\bm{x}) = \max_{l}l~f(\bm{x})$, 
we can solve problem $P$ by solving problem 
$Q$ for each $l=l,...,d_v$.
Let $\delta$ be a positive real number, and let $c_S=a_S-b_S\delta$ 
for all $S \subseteq U_i$ for $i=1,... ,d$.
\[
R : \mbox{maximize} ~~~~ 
g(\bm{x})=
  \sum_{i}\sum_{S \subseteq U_{i}} c_{S}\prod_{s \in S}x^i_{s}
~~~~ \mbox{subject to} 
~~~~ \bm{1}^T \bm{x} = l.
\]
Let $h(\delta)$ be the optimal value of problem $R$ for some $\delta$, i.e.,
$h(\delta) = \max_{\bm{x}} \{ g(\bm{x}) : \bm{1}^T \bm{x} = l \}$.
Note that we do not have to find the optimum value $\delta^{*}$ of problem $Q$,
since our goal is to determine whether $v$ maximizes its value,
that is, $l\,\delta^{*} > d_v f(\bm{1})$ or not.
All we have to do is to solve $R$ for 
$\delta' = \frac{d_v}{l}f(\bm{1})$, and 
determine whether $h(\delta') < 0$ or not. 

For converting $g(\bm{x})$ in problem $R$ to linear function,
let $y_{i,t}$ denote the 0/1 variable and let $e_{i, t}$ be the
constant for $1 \leq i \leq d$ and $1\leq t \leq |U_i|$ such that,
\begin{eqnarray*}
y_{i,t} = 
\left\{
\begin{array}{ll}
 1 & ~~~~\sum_{s \in U_i} x^i_s=t \\
 0 & ~~~~\mbox{otherwise}
\end{array}
\right. 
,~~~~~
e_{i,t}=
  \max_{S} \left\{ \sum_{T \subseteq S} c_{T} :
           S \subseteq U_{i}, |S|=t \right\}.
\end{eqnarray*}
$y_{i, t}$ indicates whether the number of edges used
in the strategy $\bm{x}$ going from $v$ to $U_{i}$ 
is equal to $t$ or not.
If $y_{i,t} = 1$, let $S \subseteq U_{i}$ be the $t$ edges chosen.
Then we consider that $g(\bm{x})$ earns $\sum_{T \subseteq S} c_{T}$,
with cost $t$. In the optimal strategy $\bm{x}$, if $y_{i,t} = 1$ 
for some $i, t$ then it must be that $S$ maximizes 
$\sum_{T \subseteq S} c_{T}$, that is $e_{i,t}$, as any other assignment
can be improved to it.
We then have the equivalent integer linear program to $R$:
\begin{align}
R' : \mbox{maximize} ~~ 
\sum^d_{i=1} ~\sum^{|U_i|}_{t=1} e_{i,t} ~ y_{i,t}
~~~
\mbox{subject to} ~~
& \sum^d_{i=1}~\sum^{|U_i|}_{t=1} t~y_{i,t} = l \\
& \sum^{|U_i|}_{t=1} y_{i,t} \leq 1 ~~ \mbox{for}~i\!=\! 1,...,d \label{difcon}.
\end{align}
Problem $R'$ is similar to a knapsack problem where each item 
has positive integer weight $t$ and value $e_{i,t}$,
and the total weight must be $l$.
The only difference is the constraint (\ref{difcon}).

Dynamic programming can be used to solve $R'$ in $O(ld_v)$ time.
Let $w(i,t)$, for $1\leq i \leq d$  and $1\leq t \leq l$, denote the 
maximum value which has total weight $t$ and uses only the $i$ first items.
Let $e_{i,0}=0$, then,
\[
w(i,t) = \max_{0 \leq s \leq |U_i|} \{ w(i-1,t-s) + e_{i,s}\}.
\]
For each $i=1,..., d$, we can compute $w(i,t)$
for $1\leq t \leq l$ in $l |U_i|$ time. 
Since $d_v = \sum_{1\le i \le d} |U_i|$,
the computation time for solving $R'$ is $O(ld_v)$.

In order to determine whether $v$ is in best response, 
we test $g(\bm{x})$ for $\delta = \frac{d_v}{1}\delta',
\frac{d_v}{2}\delta', ..., \frac{d_v}{d_v}\delta'$. 
Since $d_v \leq kd$ the running time per vertex is 
$O(2^{k}n^{3}+k^{2}d^{3}+2^{k}d)$.
Moreover $\sum_{v \in V}d_v < 2kn$ from the assumption.
Therefore, in order to determine whether every node is in best response,
it takes $O(2^{k}n^{4}+k^{2}n^{3}+2^{k}n) = O(2^{k}n^{4})$ time.
This completes the proof of Theorem \ref{genthm}.

\section{Request-Delete Model}
\label{edgedeletion2}

In this section, we study the request-delete-model for undirected PageRank games,
where not only a player $v$ can unilaterally delete bidirected links,
but also $v$ can create outlinks from $v$ to any non-neighbours. 
We may consider these outlinks to be a form of requests to the other node to
establish a link.
We consider the problem of determining whether the given input
is a Nash equilibrium in the request-delete-model or not.
We give an $O(n^{3})$ time algorithm for this 
when the underlying graph $G$ is a bidirected tree.

Let $G$ be a tree and $v$ be a node in $G$.
Let $\Gamma(v)$ denote the set of neighbours of $v$
and let $d_v=|\Gamma(v)|$.
For $i \in \Gamma(v)$, let $N_i$ be the set of nodes
which are descendants of $i$ (including $i$ itself)
in the subtree of $G$ rooted at $v$.
In the request-delete-model, a player $v$ is in best response if any combination 
of edge deletions and creation of outlinks cannot
increase PageRank of $v$.
A Nash equilibrium is a graph for which every player is in best response.
A strategy for $v$ is to retain a subset $E_v \subseteq \Gamma(v)$
of neighbours
and to choose a subset $F_v \subseteq V-\Gamma(v)$
for outlinks to her non-neighbours.
Let $\bm{x}$ and $\bm{y}$ be a $0/1$ vectors of length $d_v = |\Gamma(v)|$
and of length $|V - \Gamma(v)| = n - d_v$ respectively
which indicate $v$'s strategy.
Formally, for $i \in \Gamma(v)$, $x_i=1$ if $i$ is in $E_v$ otherwise $x_i=0$.
Similarly, for $u \in V - \Gamma(v)$, $y_u=1$ if $u$ is in $F_v$ otherwise $y_u =0$.
Let $\pi_{v}(\bm{x}, \bm{y})$ be the PageRank of $v$ on the resulting graph
for the strategy $E_v$ and $F_v$.
A node $v$ is in best response if $\pi_{v}(\bm{1}, \bm{0}) \geq \pi_{v}(\bm{x}, \bm{y})$
for any $0/1$ vectors $\bm{x}$ and $\bm{y}$.
Our approach to solve the problem for verifying a Nash equilibrium 
is to compute
the maximum of $\pi_{v}(\bm{x}, \bm{y})$ over all $0/1$
vectors $\bm{x}, \bm{y}$, for each vertex $v$.

Since $G$ is a tree and the outlinks of $v$ do not affect $\phi_{uv}$ for any $u\in V$,
equation (\ref{phitree}) holds. 
By equation (\ref{pi}), the PageRank of $v$ can be written as:
\begin{eqnarray}
\label{piout}
\pi_{v}(\bm{x}, \bm{y})=
(\bm{1}^T \bm{x} + \bm{1}^{T}\bm{y})
\dfrac{\sum_{i\in \Gamma(v)}a_ix_i}
{\sum_{i\in \Gamma(v)}
   \left(\left(1-c_i\right)x_i
     +\sum_{u\in N(i)-\{i\}}\left(1-e_u x_i\right)y_u\right)}.~~
\end{eqnarray}
where we let $\bm{a}=(a_i)$, $\bm{c}=(c_i)$ and $\bm{e}=(e_u)$
be constants such that
\begin{eqnarray*}
a_i&=&\alpha \sum_{u \in N_i}q_{u}\phi_{uv},
~~ c_i = (1-\alpha)\phi_{iv} ~~~~\mbox{for}~i \in \Gamma(v). \\
e_u&=&(1-\alpha)\phi_{uv} ~~~~\mbox{for}~u \in V-\Gamma(v).
\end{eqnarray*}

The proof of the following lemma can be found in \cite{bib1}.
\begin{lemma}\cite{bib1}
\label{lemnbr}
Let $D$ be a directed graph.
For a node $v$ in $D$, if a node $u \neq v$ has the maximum potential
with respect to $v$ then $u$ is a in-neighbour of $v$.
\end{lemma}
We prove following two lemmas.
\begin{lemma}
\label{lembc}
Let $G$ be a tree, $v$ be a node in $G$, 
and let $N_i$ be defined as above.
For all $i \in \Gamma(v)$ and for all $u \in N(i)-\{i\}$,
\[
\phi_{iv} > \phi_{uv}.
\]
\end{lemma}
{\it Proof.}
Let $T$ be the subtree of $G$ induced by $v$ and all nodes in $N_i$.
Since $G$ is a tree, 
the walk starting at the nodes in $N_i$ cannot
reach the nodes in $V-N_i$ without visiting $v$.
Thus the potential from each vertex $u \in T$ with respect to $v$ is
the same as the potential in $G$.
By Lemma \ref{lemnbr}, only a neighbour of $v$ can have the 
maximum potential to $v$ in $T$
and the other potentials are strictly less.
This means $\phi_{iv} > \phi_{uv}$ for all $u \in N(i)$ in $G$
and concludes the proof.
\qed

By Lemma \ref{lembc}, we have the following strict inequality.
\begin{equation}
\label{eqbc}
c_i > e_u~~~~\mbox{for $u \in N_i$, for $i \in \Gamma(v)$}.
\end{equation}

Let $\hat{\bm{y}}^{l}$ denote a $0/1$ vector over $V-\Gamma(v)$ such that
$\hat{y}^{l}_{u} = 1$ if $e_u$ is
within the $l$th largest values in the all entries of $\bm{e}$,
and $\hat{y}^{l}_{u} = 0$ otherwise.
\begin{lemma}
\label{lemhy}
Let $\pi_v(\bm{x},\bm{y})$ and $\hat{\bm{y}}^{l}$ be defined as above.
For $l = 1,...,n$,
\begin{eqnarray*}
&&\max_{\bm{x}, \bm{y}}~
\{ ~\pi_v(\bm{x},\bm{y})
   ~|~\bm{1}^{T}\bm{x} + \bm{1}^{T}\bm{y} = l ~\} \\
&&=
\max_{\substack{ \bm{x} \\ 1 \leq l_1 \leq d_v \\~~~0\leq l_2\leq n-d_v}}
\{ ~\pi_v(\bm{x},\hat{\bm{y}}^{l_2})
   ~|~\bm{1}^{T}\bm{x}=l_1,~ l_1+l_2=l~\}.
\end{eqnarray*}
\end{lemma}
{\it Proof.}
By contradiction.
When $\bm{y} = \bm{0}$ and $\bm{y} = \bm{1}$,
obviously the above equation holds,
so we consider the case of $\bm{y} \neq \bm{0}, \bm{1}$.

Assume that $(\bm{x}^{*}, \bm{y}^{*})$ is the maximum assignment
for $\pi_v(\bm{x},\bm{y})$ subject to
$\bm{1}^{T}\bm{x} + \bm{1}^{T}\bm{y} = l$,
and that $\bm{y}^{*} \neq \hat{\bm{y}}^{l_2}$ for any
$1 \leq l_2 \leq n-d_v-1$.
There exist $u, w$ in $V-\Gamma(v)$
such that $e_{u} > e_{w}$ and $y^{*}_u=0$, $y^{*}_{w}=1$.
Let $i$ denote the neighbour of $v$ such that $u$ is in $N_i$.

When the case of $x^{*}_i=1$,
the assignment $y^{*}_u=1$, $y^{*}_{w}=0$
decreases the denominator in the equation (\ref{piout}),
since $1-e_u < 1-e_w$.
This gives an improvement for $\pi_v(\bm{x},\bm{y})$,
and does not change the Hamming weights of $\bm{x}$ and $\bm{y}$,
a contradiction.
Therefore we take $x^{*}_i=0$. However the assignment
$x^{*}_i=1$, $y^{*}_{w}=0$ 
decreases the denominator in the equation (\ref{piout}),
since $1-c_i < 1-e_u < 1-e_w$ by (\ref{eqbc}).
This gives an improvement,
and does not change the value $\bm{1}^{T}\bm{x} + \bm{1}^{T}\bm{y}$.
This contradiction concludes the proof.
\qed

Lemma \ref{lemhy} means that if a player $v$ is in best response,
$v$ puts her outlinks to the nodes which have the higher potential to $v$.

\begin{theorem}
\label{thmout}
Given a bidirected tree $G$, jump probability $\alpha$ and distribution
$\bm{q}$, we can determine in $O(n^3)$ time whether this
is a Nash equilibrium in the request-delete-model and if not give an improving strategy
for at least one player.
\end{theorem}
{\it Proof.}
Consider the following fractional integer programming problem:
\begin{eqnarray*}
P' : \mbox{maximize} ~~~~ \pi_{v}(\bm{x}, \bm{y})
~~~~ \bm{x} \in \{ 0,1 \}^{d_v},
~~ \bm{y} \in \{ 0,1 \}^{n-d_v}.
\end{eqnarray*}
We solve optimization problem $P'$ for each $v$ in $V$.
For given $v$ and each $l_2 = 0,...,n-d_v$,
we compute $\hat{\bm{y}}^{l_2}$ as defined just before Lemma \ref{lemhy}.
Consider the following $0/1$ optimization problem $Q'$
for $l_1 = 1,...,d_v$.
\[
Q' : \mbox{maximize} ~~~~
f(\bm{x})=
~
\frac{\bm{a}^T \bm{x}}
{\bm{b}^T \bm{x}}.
~~~~\mbox{subject to}~~
\bm{1}^{T}\bm{x} = l_1
\]
where $b_i =
((1-c_i)
  +\sum_{u\in N(i)-\{i\}}(1-e_u)\hat{y}^{l_2}_u)$ for $i \in \Gamma(v)$.
By using the algorithm to solve the problem $Q$ in Section \ref{trees},
we can solve problem $Q'$
for all $l_1 = 1,...,d_v$ in $O(d_v^{2})$ time.
Let $l = l_1 + l_2$.
By Lemma \ref{lemhy},
the solution to problem $Q'$ for each $l = 1,...,n$ gives also the
solution to $P'$ since $\pi_{v}(\bm{x}, \hat{\bm{y}}^{l_2}) = (l_1 + l_2)f(\bm{x})$.
Thus, for each node $v$ in $V$, we can determine 
whether $v$ is in best response.
It follows that we can determine whether the input is a Nash equilibrium.

Finally, we analyze the running time of the algorithm.
It takes $O(n^{2})$ time to compute all potentials in $G$.
For each $l_2=0,1,...,n-d_v$, it takes $O(d_v^{2})$ times to solve
the problem $Q'$.
Therefore we can determine if a node $v$ is in best response
in $O\left( \left( n-d_v \right) d_{v}^{2} \right)$ time.
Since $G$ is a tree, $\sum_{v\in V} d_v = O(n)$.
Thus we can test whether all nodes are in best response
in $O(n^3)$ time.
\qed

\section{Add-Delete Model}
\label{adddel}

In this section, we introduce the add-delete-model,
where each player $v$ 
can delete any edges from $v$ 
and can add one edge to any non-neighbour $u$ 
if by so doing the PageRank of $u$ increases. Otherwise we
may presume that $u$ would simply delete the edge $(u,v)$.
We consider the problem of determining whether or not the input
is a Nash equilibrium and
give an $O(2^{2k} n^5)$ time algorithm for general graphs,
where $k$ is the maximum vertex degree on any biconnected 
component in the graph.

Let $G=(V,E)$ be an undirected graph and let $k=k(G)$
be a parameter defined as follows.
If $G=(V,E)$ is a tree we set $k=1$ else $k$ is the maximum
vertex degree in any biconnected component of $G$.
Let $\Gamma(v)$ denote the set of neighbors
of a node $v$ in $G$. Let $d_v = |\Gamma(v)|$.
In the add-delete-model, the strategy for a player $v$ is 
to retain the subset of neighbours from $\Gamma(v)$ 
and to choose
one non-neighbour $u$ to add an edge between $u$ and $v$.
The PageRank of $u$ must increase by the strategy.
A player $v$ is in best response if $v$ cannot increase
her PageRank by her any other possible strategies.
A Nash equilibrium is a graph where every player is in best response.

Let $v$ be a node in $G$ and
let $\{ C_1, C_2, ..., C_d \}$ be the set of connected components
in the subgraph that are induced by deletion of $v$ from $G$.
It follows from the definition of $k=k(G)$ that
$v$ has at most $k$ links to $C_{i}$ for $i=1,..., d$.
Let $U_{i} = \{ u : u \in C_{i}, u \in \Gamma(v) \}$ be
the set of $v$'s neighbours in $C_i$ for $i=1,...,d$.
Let $\bm{x}$ be a $0/1$ vector of length $|\Gamma(v)|$
which indicates the strategy of $v$ for edge deletion.
We write $\bm{x}=(x^1,x^2,...,x^d)$ as the concatenation of the 0/1
vectors $x^i$ representing the strategy
restricted to the component $C_i$ for $i=1,...,d$.
The potential $\phi_{wv}(\bm{x})$ on $G$ is given by equation 
(\ref{phieq}) for each $w$ in $V$.

\begin{theorem}
\label{adddelthm}
Given a graph $G$ with $k=k(G)$, jump probability
$\alpha$ and distribution $\bm{q}$, 
in $O(2^{2k} n^5)$ time we can
determine if this is a Nash Equilibrium in the add-delete-model 
and if not
give an improving strategy for at least one player. 
\end{theorem}
{\it Proof.}
We give an algorithm which determines whether a player $v$ 
in best response, for each vertex $v$ in $V$.

We fix $v$ and for each of $v's$ non-neighbours $u$
perform the following steps.
Let $C_j$ be the component containing $u$.
We initially decide the strategy vector $x^{j}$ for $C_j$ by choosing
a subset $U'_j \subseteq U_j$.
Let $G'$ be the graph formed by adding $(u,v)$ to $G$ and 
by deleting the edges to nodes not in $U'_j$.
Otherwise we
retain the edges to the nodes in $U_i$, for each $i = 1,...d$, $i \neq j$.
Let $\bm{x}^{-j} = (x^1,...,x^{j-1},x^{j+1},...,x^d)$.
For each node $w$ in $C_j$, 
let $\phi^{G'}_{wv}$ be the potential calculated on $G'$.
Since $v$ is a cut vertex of $G'$,
these potentials are invariant to $\bm{x}^{-j}$.
By equation (\ref{pi}), $v$'s PageRank on the resulting graph for the 
strategy $\bm{x}^{-j}$ on $G'$ can be written as follows:
\begin{equation}
\pi_{v}^{G'}(\bm{x}^{-j}) = 
  (\bm{1}^T\bm{x}^{-j} + |U'_{j}|)
    \dfrac{a_0 +
          \sum_{i \neq j}\sum_{S \subseteq U_{i}}a_{S}\prod_{s \in S}x^i_{s}}
          {b_0 +
          \sum_{i \neq j}\sum_{S \subseteq U_{i}}b_{S}\prod_{s \in S}x^i_{s}}
\end{equation}
where $a_{S}$, $b_{S}$ are defined 
by equation (\ref{generala}) and (\ref{generalb}),
and $a_{0} = \alpha \sum_{w \in C_j}q_w \phi^{G'}_{wv}$ 
and $b_{0} = |U'_{j}| - (1-\alpha)\sum_{w \in U'_j}\phi^{G'}_{wv}$
are known constants.
We consider the following problem.
\begin{equation*}
P'':\mbox{maximize} ~~~~
\pi_v^{G'}(\bm{x}^{-j})
~~~~
\mbox{subject to}~~~~
\pi_u^{G'}(\bm{x^{-j}}) > \pi_u^{G}
\end{equation*}
In order to determine whether $v$ maximizes its PageRank, 
we fix the Hamming weight $l$ of $\bm{x}^{-j}$
for each $l = 1,...,|\Gamma(v)| - |U'_j|$
and consider the following problem $Q''$.
\begin{eqnarray}
Q'': \mbox{maximize} ~
 \dfrac{a_0 +
       \sum_{i\neq j}\sum_{S \subseteq U_{i}}
           a_{S}\prod_{s \in S}x^i_s}
       {b_0 +
       \sum_{i\neq j}\sum_{S \subseteq U_{i}}
           b_{S}\prod_{s \in S}x^i_s}
~~
\mbox{subject to}~~
\bm{1}^T \bm{x}^{-j} &=& l \\
\pi_u^{G'}(\bm{x^{-j}}) &>& \pi^{G}_u. \label{const}~~~~
\end{eqnarray}

According to the PageRank formulae for each $w$ in $C_j$,
\[
\pi_w^{G'}(\bm{x}^{-j}) =
  \alpha q_w + 
  \sum_{w' \in \Gamma(w)} \frac{\pi_{w'}^{G'}(\bm{x}^{-j})}
                               {|\Gamma(w')|}.
\]
Since the nodes in $C_j$ are not adjacent to the nodes in $C_i$ 
for each $i=1,...,d$, $i \neq j$, we can calculate the PageRank of $u$ as follows.
\begin{equation*}
\pi_u^{G'}(\bm{x}^{-j}) =
 \zeta_u + \eta_u\frac{\pi_v^{G'}(\bm{x}^{-j})}{\bm{1}^{T}\bm{x}^{-j} + |U'_{v}|}
\end{equation*}
where $\zeta_u$ and $\eta_u$ are positive constants.
Thus the constraint (\ref{const}) in the problem $Q''$
is $\pi_{v}^{G'}(\bm{x}^{-j}) > \frac{l+|U'_v|}{\eta_u}(\pi_u^{G} - \zeta_u)$.
Since our goal is to maximize $\pi_{v}^{G'}(\bm{x}^{-j})$,
we can ignore the constraint (\ref{const}) during the maximization.
Let $\delta^{*}$ be the optimal value of problem $Q''$ without
the constraint (\ref{const}).
Player $v$ is not in best response if
$\pi_v^{G}/(l+|U'_v|) < \delta^{*}$ 
and 
$\frac{l+|U'_v|}{\eta_u}(\pi_u^{G} - \zeta_u) \leq \delta^{*}$.
We can test these inequalities by the algorithm for problem $Q$
in Section \ref{general}.
This means that we can determine if $v$ is in
best response or not.

Finally, we analyze the running time of the algorithm.
For each non-neighbour $u$ of $v$ and 
for each subset $U'_j$ of $U_j$,
it takes $O(n^{3})$ time to compute the potential $\phi^{G'}_{wv}$ 
for each $w$ in $C_j$.
It takes $O(2^{k}n^{3})$ time for problem $Q''$ for each
$l = 1,...,|\Gamma(v)| - |U'_j|$.
Therefore it takes $O(2^{2k}n^{4})$ time per vertex and
so we can determine if the input is a Nash equilibrium or not in 
$O(2^{2k}n^{5})$ time.
\qed

\subsection{$\alpha$-insensitive equilibria}
We prove that $G$ is an $\alpha$-insensitive equilibrium in the add-delete-model
if and only if $G$ is a complete graph using the following lemma.

\begin{lemma}
\label{zeroalpha}
If an undirected graph $G$ is connected,
for non-neighbours $u,v$ in $V$ there exists $\alpha_{0}$
such that for all $0 < \alpha \leq \alpha_0$,
\begin{eqnarray*}
\pi_{v}^{G'} > \pi_{v}^{G},~~ \pi_{u}^{G'} \geq \pi_{u}^{G}
\end{eqnarray*}
where $G' = \left(V, E \cup \left(u,v \right) \right)$.
\end{lemma}
{\it Proof.}
Recall that the stationary distribution in a standard random walk 
on a connected undirected graph is proportional to the degree of each vertex.
In an $\alpha$-random walk,
when we have $\alpha \rightarrow 0$, $\pi_v^{G}$ converges to $d_v/d'$
and $\pi_v^{G'}$ converges to $(d_v+1)/(d'+2) > d_v/d'$,
where $d' = \sum_{v\in V} d_v$ and so $d' > 2d_v$.
By the continuity of $\pi_v$ for $\alpha \in (0,1)$ \cite{chen}, 
there exists $\alpha_0$ such that
$\pi_{v}^{G'} > \pi_{v}^{G}$ and $\pi_{u}^{G'} > \pi_{u}^{G}$
for all $0 < \alpha \leq \alpha_0$.
This concludes the proof.
\qed

An undirected graph $G$ is called an $\alpha$-insensitive equilibrium
if $G$ is an equilibrium for all $0 < \alpha < 1$.

\begin{theorem}
\label{thmcomplete}
If an undirected graph $G$ is connected,
$G$ is an $\alpha$-insensitive equilibrium
in the add-delete-model
if and only if $G$ is a complete graph.
\end{theorem}
{\it Proof.}
Assume that $G$ is an $\alpha$-insensitive equilibrium
and is not a complete graph, and thus $G$ has at least one non-neighbours
of $(u,v)$.
By Lemma \ref{zeroalpha}, there exists $\alpha_0$ such that
for $\alpha < \alpha_0$, the edge addition $(u,v)$ must increase
the PageRank of both $\pi_u$ and $\pi_v$.
This contradicts that $G$ is an equilibrium.

Conversely, assume that $G$ is a complete graph.
There does not exist a non-neighbour in $G$.
$G$ is a Nash equilibrium
in the directed PageRank game since each node is in best response (see Section \ref{DPG}).
Therefore any deletion of undirected edges for a node $v$ does not
improve her PageRank for any $\alpha$.
This means $G$ is an $\alpha$-insensitive equilibrium
in the add-delete-model.
\qed

One may wonder if there are $\alpha$-sensitive equilibrium
in the add-delete-model which are connected and are not complete graphs.
We do not know the answer to this question even if we assume
$q$ is a uniform distribution.
However, we have found
some $\alpha$-sensitive equilibria in the add-delete-model \emph{without}
deletions allowed. That is, we allow single edge additions 
that raise the PageRank of both endpoints but we do not allow deletions.
In the case of non-uniform $q$ we have found many 
examples of $\alpha$-sensitive equilibria. For uniform $q$ we have the following
examples.

\begin{figure}[htbp]
\begin{center}
\begin{tabular}{lcr}
  \begin{minipage}{0.315\hsize}
    \begin{center}
    \includegraphics[width = \hsize]{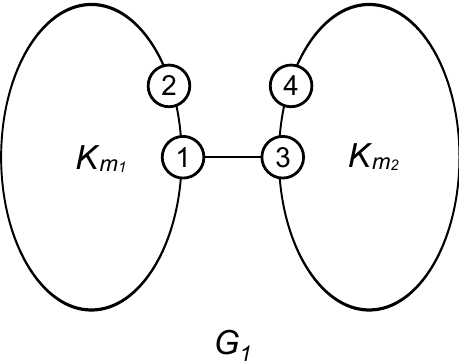}
    \end{center}
  \end{minipage}
  ~~~~~~
  \begin{minipage}{0.50\hsize}
    \begin{center}
    \includegraphics[width = \hsize]{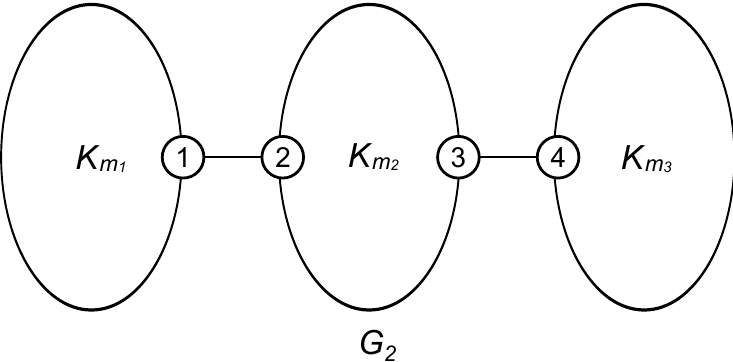}
    \end{center}
  \end{minipage}
\end{tabular}
\caption{$\alpha$-sensitive equilibria: add-delete-without deletions, uniform $q$}
\label{alphasensitive}
\end{center}
\end{figure} 

Figure \ref{alphasensitive}
shows two graphs $G_1$ and $G_2$
which are examples of
$\alpha$-sensitive equilibria in
the no-deletion add-delete-model with uniform $q$, where $K_m$ denotes the complete graph of size $m$.
In $G_1$, there are three possible choices for an edge addition which are:
(i) 1,3, (ii) 1,4 and (iii) 2,4.
For example, take $m_1 = 50$, $m_2 = 2$ and $\alpha = 0.15$.
Case (i) decreases PageRank of node 1,
case (ii) decreases PageRank of node 1
and case (iii) decreases PageRank of node 2.
Therefore $G_1$ is an equilibrium for $\alpha = 0.15$.
While we take $\alpha = 0.02$,
any of these cases increase the PageRanks of the two endpoints.
Therefore $G_1$ is not an equilibrium for $\alpha = 0.02$.
It follows that $G_1$ is an $\alpha$-sensitive equilibrium.
The graph $G_2$ is also an $\alpha$-sensitive equilibrium
for $m_1=10000$, $m_2=100$, $m_3=2$ and $\alpha = 0.15$.
We remark that these are not equilibria if deletions are allowed.
In $G_1$ the deletion of $(1,3)$ improves the PageRank of node $3$.
Similarly in $G_3$, the deletion of $(3,4)$ improves the PageRank of
node $4$. 

\subsection{A Sufficient Condition for Edge Addition}

Throughout this section,
we assume that the distribution vector $\bm{q}$ is uniform,
i.e., $q_v = 1/n$ for all $v \in V$.
We show that if $G$ is symmetric with respect to
non-neighbours $u,v$ in $G$, then an edge addition
between $u,v$ must increase the PageRank of both $u$ and $v$.
An automorphism mapping on $G$ is a permutation
$\sigma$ over the vertices
such that $(\sigma(u),\sigma(v)) \in E$ if and only if $(u,v) \in E$.
Let Aut$(G)$ be a set of automorphisms on $G$.

\begin{theorem}
If an undirected graph $G$ has non-neighbours $u,v$
and has a graph automorphism
$\sigma \in \mbox{Aut}(G)$ such that $\sigma(v) = u$
and $\sigma(u) = v$, then
\begin{eqnarray*}
\pi_{v}^{G'} > \pi_{v}^{G},~~ \pi_{u}^{G'} > \pi_{u}^{G}
\end{eqnarray*}
where $G' = \left(V, E \cup \left(u,v \right) \right)$.
\end{theorem}
{\it Proof.}
Let $z_{ij}$ be the expected number of times
in which an $\alpha$-random walk visits $j$ before the first jump 
when the walk starts at $i$.
\[
z_{ij}
=E \left[ \left. 
      \sum_{t=0}^{J-1}I\left\{X_t=j\right\}
          \right| X_0=i
   \right]
\]
where $J \geq 1$ is the time of the first jump and $I\{X_t=j\}$ denotes
the indicator for the $t$th step of $\alpha$-random walk visiting $j$.
Since the mapping $\sigma$ preserves adjacency of all neighbours in $G$,
we have $z_{ij} = z_{\sigma(i)\sigma(j)}$ for all $i,j \in V$.
Since $\bm{q}$ is the uniform distribution,
\begin{eqnarray*}
\pi_{v}^{G}
&=&\frac{\alpha}{n}\sum_{w \in V}z_{w,v} \\
&=&\frac{\alpha}{n}\sum_{w \in V}z_{\sigma(w),\sigma(v)} \\
&=&\frac{\alpha}{n}\sum_{w \in \sigma(V)}z_{w, u} \\
&=&\frac{\alpha}{n}\sum_{w \in V}z_{w,u} = \pi_{u}^{G}.
\end{eqnarray*}
Similarly, $\left(\sigma\left(u\right), \sigma\left(v\right)\right) = (v,u)$
is in $G'$, so the mapping
$\sigma$ is in Aut($G'$).
Hence we have $\pi_{v}^{G'} = \pi_{u}^{G'}$.

For a node $w$ in $V$, let $z_{w[uv]}$ denote the expected number of times
in which the walk visits $u$ and $v$ when the walk starts at $w$ before a random jump.
By the linearity of expectations,
\begin{eqnarray*}
z_{w[uv]}
&=&E \left[ \left. \sum_{t=0}^{J-1}
\left(I\left\{X_t=u\right\}+I\left\{X_t=v\right\}\right)
\right| X_0=w \right] \\
&=&E \left[ \left. \sum_{t=0}^{J-1}I\{X_t=u\} \right| X_0=w \right] +
   E \left[ \left. \sum_{t=0}^{J-1}I\{X_t=v\} \right| X_0=w \right] \\
&=&z_{wu} + z_{wv}.
\end{eqnarray*}
Let $\pi_{[uv]}^{G}$ denote the sum of PageRank of $u$ and $v$ on $G$.
Then we have,
\begin{eqnarray*}
\pi_{[uv]}^{G}
&=&\pi_{v}^{G} + \pi_{u}^{G} \\
&=&\frac{\alpha}{n}\sum_{w \in V} (z_{wu} + z_{wv})
=\frac{\alpha}{n}\sum_{w \in V} z_{w[uv]}. \label{piuandv}
\end{eqnarray*}

We compute the expected number of visits to $u$ and $v$
when the walk starts at $w$.
Firstly we compute the probability that the walk reaches $u$ or $v$ and
consider the expected number of walks revisiting $u$ and $v$.
We have $z_{w[uv]} = \phi_{w[uv]} z_{[uv][uv]}$,
where $\phi_{w[uv]}$ is the probability that a walk starting from $w$
reaches $u$ or $v$ before the first jump.
The term $z_{[uv][uv]}$ counts the 
number of returns from $u \cup v$
to $u \cup v$ before the first jump occurs.
\[
z_{[uv][uv]}
= E \left[\left.\sum_{t=0}^{J-1}
    I\{X_t=u \cup X_t=v\} \right| X_0=u \cup X_0=v \right].
\]
Each return succeeds with probability $\phi^{+}_{[uv][uv]}$,
where $\phi^{+}_{[uv][uv]}$ is the probability of the walk 
returning from $u \cup v$
to $u \cup v$ before the first jump.
Hence we have
\[
z_{[uv][uv]}
= 1 + \phi^{+}_{[uv][uv]} + (\phi^{+}_{[uv][uv]})^{2} + \cdots
= \frac{1}{1-\phi^{+}_{[uv][uv]}}.
\]
Since we have $\sigma(v) = u$, $\sigma(u) = v$ and $q_v = q_u = 1/n$,

\begin{eqnarray}
\label{phiplusuandv}
\phi^{+}_{[uv][uv]} = \phi^{+}_{u[uv]} = \phi^{+}_{v[uv]}
= \frac{(1-\alpha)}{|\Gamma(v)|}
\sum_{i \in \Gamma(v)} \phi_{i[uv]}.
\end{eqnarray}

Now we conclude the theorem.
If we add a new edge between $u$ and $v$ to $G$,
the right-hand in (\ref{phiplusuandv}) becomes strictly
larger than the before, because we have
$\phi_{v[uv]} = 1$.
By equation(\ref{piuandv}),
this means that $\pi_{[uv]}^{G'} > \pi_{[uv]}^{G}$.
We also have $\pi_{[uv]} = \pi_{u} + \pi_{v} = 2 \pi_{u} = 2 \pi_{v}$.
This concludes the proof.
\qed

\section{Conclusion}
We have constructed three different models for PageRank games
on undirected web graphs and
studied the problem of verifying Nash equilibria
in each model.
The algorithms obtained have high complexity and are thus useable only on
small scale networks, or on a small fragment of a much larger network as an
approximation. From the algorithmic side, the main open question is whether
polynomial time algorithms exist for checking Nash equilibria in general graphs.

Other results in this paper concern the evolution of PageRank graphs if players
act unilaterally to improve their page rank. We showed in Theorem \ref{thmcomplete} that
complete graphs are the only $\alpha$-insensitive equilibria in the add-delete-model.
It would clearly spell doom for any social network if this happened!
Are there any non-$\alpha$-insensitive equilibria?

\subsubsection*{Acknowledgements.}
We thank Yuichi Yoshida and Junichi Teruyama for many helpful discussions.

\end{document}